\documentclass[10pt, letterpaper]{article}

\usepackage[OT1]{fontenc}
\usepackage[in]{fullpage}
\usepackage{titlesec}

\usepackage[pdftex,pdfpagemode={UseOutlines},bookmarks,bookmarksopen,colorlinks,linkcolor={black},citecolor={black},urlcolor={red}]{hyperref}
\usepackage{graphicx}
\usepackage{caption}
\usepackage{subcaption}
\usepackage{bm}
\usepackage{amsmath,amssymb}
\usepackage{aas_macros}
\usepackage{float}

\graphicspath{{figure/}}

\usepackage{mathptmx}  

\titleformat*{\section}{\normalsize\bfseries\scshape}
\titleformat*{\subsection}{\normalsize\bfseries\scshape}
\titleformat*{\subsubsection}{\normalsize\bfseries\scshape}
\titleformat*{\paragraph}{\normalsize\bfseries\scshape}
\titleformat*{\subparagraph}{\normalsize\bfseries\scshape}

\newcommand{\half}{\frac{1}{2}}

\newcommand{\fc}{\mathbf{H}}
\newcommand{\fcx}{H_x}
\newcommand{\fcy}{H_y}
\newcommand{\psf}{\Pi}
\newcommand{\psfwidth}{\sigma_{\psf}^{2}}

\newcommand{\erf}{\rm Erf}
\newcommand{\modelparams}{\omega}
\newcommand{\modelparamsnoflux}{\Omega}

\begin{document}
  
\title{\large{{\bf Blind Detection of Ultra-faint Streaks with a Maximum Likelihood Method}}}

\author{
\large{{\bf William A. Dawson, Michael D. Schneider, \& Chandrika Kamath}}\\
\large{{\it Lawrence Livermore National Laboratory, 
P.O. Box 808 L-211, Livermore, CA 94551-0808, USA.}}\\
\vspace{0.01in}\\
\normalsize{LLNL-CONF-703048}
}

\date{}

\maketitle

\section{Abstract}  

We have developed a maximum likelihood source detection method capable of
detecting ultra-faint streaks with surface brightnesses approximately an order
of magnitude fainter than the pixel level noise. Our maximum likelihood
detection method is a model based approach that requires no {\it a priori}
knowledge about the streak location, orientation, length, or surface brightness.
This method enables discovery of typically undiscovered objects, and enables the
utilization of low-cost sensors (i.e., higher-noise data). The method also
easily facilitates multi-epoch co-addition. We will present the results from the
application of this method to simulations, as well as real low earth orbit
observations.

\section{Introduction}
\label{sec:introduction}

Satellites and near earth objects (NEOs) appear as streaks in the images of
sidereal tracking surveys.  There is evidence that the NEOs and satellites
(including space debris) follow an inverse power law luminosity function with
there being many more fainter objects than brighter objects
\cite{Schildknecht:2001ks, Schildknecht:2004kr, NAP12842}. Thus going fainter
provides the best return on object characterization and actionable information.
Additionally, algorithms that can detect streaks as faint as possible can enable
the use of lower cost telescopes and detectors.

Astronomical surveys of local solar system or Earth orbiting objects typically
fall into two broad categories: object tracking and sidereal tracking. In the
case of object tracking, the object of interest will appear as a point source
convolved with the atmospheric/optic point spread function (PSF) and the stars
or other objects moving at a different angular velocity will appear as streaks.
In the case of sidereal tracking, the stars will appear as PSF's and satellites
and NEOs will appear as streaks of varying lengths. Sidereal tracking surveys
have the advantage of enabling easy discrimination of satellites and NEOs from
galactic and extra-galactic sources, which significantly outnumber the
satellites and NEOs.  This comes at the price of making detection of satellites
and NEOs more challenging due to extending their signal spatially and increasing
the noise floor as a function of the angular velocity of the source
\cite{Krugly:2004ca}. We will introduce a maximum likelihood detection method as
a statistically rigorous extension of a signal-matched-filter
\cite{woodward1953probability,turin1960} that is capable of maximizing the
detectability of satellites and NEOs in both categories of surveys while
maintaining high purity.


\section{Method}
\label{sec:method}

In general, a signal-matched-filter will maximize the detectability of a given
object. In the case of a ground-based object tracking survey the signal matched
filter is the PSF, and in the case of a sidereal tracking survey it is a finite
line convolved with the PSF, see for example the work by \cite{SchneiderDawson}
presented at this conference or \cite{Veres:2012hu,
Kouprianov:2008gs,Schildknecht2015}. In addition to incorporating a signal-
matched-filter, methods that can account for spatially and temporally varying
pixel covariance, facilitate the combination of multiple epochs of images
\cite{Szalay1999}, incorporate prior information, and account for pixel level
systematics (e.g., aliasing), among other basic image reduction tasks (e.g.,
flat-fielding, background subtraction, etc.), can further increase the magnitude
limit of detectability. The maximum likelihood detection method exemplifies all
of these beneficial features and is furthermore statistically sound.

Kaiser developed and discussed in great detail a maximum likelihood detection
method for PSF sources as part of the Pan-STARRS survey
\cite{Kaiser2001,Kaiser2002}. In this section we generalize this method to
include extended sources and specifically discuss the applicability of the
method to detecting streaks. Additionally, we discuss some of the practical
considerations in applying this method.

\subsection{General Form}

In this section we derive the general form of the maximum likelihood source
detection method. The likelihood of there being a source at location
$\mathbf{x}$ in the image plane with properties $\modelparams$ is,

\begin{equation}
\mathrm{ln}\mathcal{L}\left(\mathbf{x}, \modelparams \right) = -\tfrac{1}{2} \left[ d(\mathbf{x})-m(\mathbf{x},\modelparams ) \right]^\mathrm{T} \Sigma\left(\mathbf{x}, \modelparams \right)^{-1} \left[ d(\mathbf{x})-m(\mathbf{x},\modelparams ) \right] + C,
\end{equation}
where $d$ is the pixel data (e.g., analog-to-digital units), $m$ is the source
model that can be the convolution of the source $s$ with the PSF $\Pi$,
\begin{equation}
m (\mathbf{x}, \modelparams) = \int s(\mathbf{x},\modelparams) \Pi(\mathbf{x}) \mathrm{d}\mathbf{x},
\end{equation}
$\Sigma$ is the pixel covariance, and $C$ is a constant. We can maximize this
likelihood with respect to the object flux,
\begin{equation}
m (\mathbf{x}, \modelparams) \equiv \alpha\mu(\mathbf{x}, \modelparamsnoflux),
\end{equation}
where  $\mu$ is a source image model with unity flux, $\alpha$ is a
multiplicative scaling of the unit flux, and $\modelparamsnoflux$ is the same as
the set $\modelparams$ but excluding the flux or surface brightness parameter.
It is also convenient to expand the likelihood,
\begin{equation}
\mathrm{ln}\mathcal{L}\left(\mathbf{x}, \modelparams \right) = -\tfrac{1}{2} d^\mathrm{T} \Sigma^{-1} d -\tfrac{1}{2} m^\mathrm{T} \Sigma^{-1} m + d^\mathrm{T} \Sigma^{-1} m + C,
\end{equation}
and define the functions
\begin{equation}\label{eq:phi}
\Phi (\mathbf{x}, \modelparamsnoflux) = \mu(\mathbf{x}, \modelparamsnoflux)^\mathrm{T} \Sigma^{-1} \mu(\mathbf{x}, \modelparamsnoflux),
\end{equation}
and
\begin{equation}\label{eq:psi}
\Psi (\mathbf{x}, \modelparamsnoflux) = d(\mathbf{x})^\mathrm{T} \Sigma^{-1} \mu(\mathbf{x}, \modelparamsnoflux).
\end{equation}
These are the noise weighted model auto-correlation, and noise weighted 
data-model cross-correlation, respectively. In these terms the maximum 
likelihood can be solved for with 
\begin{equation}\label{eq:maxlike}
\frac{\partial \mathrm{ln}\mathcal{L}}{\partial \alpha} = \alpha \Phi(\mathbf{x}, \modelparamsnoflux) - \Psi(\mathbf{x}, \modelparamsnoflux) = 0.
\end{equation}
Solving \autoref{eq:maxlike} we see that the maximum likelihood flux of the 
model object at position $\mathbf{x}$ is
\begin{equation}
\alpha_\mathrm{ML} = \frac{\Psi(\mathbf{x}, \modelparamsnoflux)}{\Phi(\mathbf{x}, \modelparamsnoflux)},
\end{equation}
and the maximum likelihood is,
\begin{equation}\label{eq:loglike}
\mathrm{ln}\mathcal{L}_\mathrm{ML} = \frac{\Psi^2(\mathbf{x}, \modelparamsnoflux)}{2\Phi(\mathbf{x}, \modelparamsnoflux)}.
\end{equation}
Given this we can express the significance of there being an object at location
$\mathbf{x}$ with properties $\modelparamsnoflux$ as
\begin{equation}\label{eq:sig_def}
    \nu(\mathbf{x}, \modelparamsnoflux) = \frac{\Psi(\mathbf{x}, \modelparamsnoflux)}{\sqrt{\Phi(\mathbf{x}, \modelparamsnoflux)}}.
\end{equation}

There are a number of beneficial features associated with this method. If we
wish to combine multiple epochs we simply need to sum the log likelihoods,
\autoref{eq:loglike}, of the different epochs (see also \cite{Babu2008}). For
example, if the source moves or if the PSF changes between epochs $i$ and $j$ we
can simply sum $\mathrm{ln}\mathcal{L}_\mathrm{ML}(\mathbf{x}_i,
\Pi_i(\mathbf{x}_i), \cdot) + \mathrm{ln}\mathcal{L}_\mathrm{ML}(\mathbf{x}_j,
\Pi_j(\mathbf{x}_j), \cdot)$ and estimate a combined significance of detection.
Note that because the coordinates $\mathbf{x}$ need not be integer pixel values,
this avoids many systematics associated with working on a pixel grid and
obviates the need for complex pixel transformations before co-addition.  This
also more rigorous than many common co-addition practices, such as convolving to
the images taken with better seeing (i.e., smaller PSF) by that of the worst
seeing image's PSF before co- adding data, which throw away information. It is
also trivial to incorporate spatially varying source models, image noise, and
PSFs.

\subsection{Streak Model}

Since point source models (i.e., $s=\delta(\mathbf{x})$) are more
straightforward and have been treated in great detail by Kaiser
\cite{Kaiser2001,Kaiser2002}, we will focus on streak models for this report. As
detailed in \cite{SchneiderDawson}, we model a streak in image space as a
product of a narrow Gaussian (representing the width that can be made
arbitrarily small) and a rectangular window (representing the extent of the
streak; see also \cite{Veres:2012hu}). We further assume that the streak model
is evaluated in a coordinate system with the center of the streak at the origin
and the extent of the streak along the x-axis. This can always be accomplished
by means of a coordinate translation by $x_0$, $y_0$ and a rotation by $\phi_0$.
Then a streak  of length $L$, width $\sigma_y$, and surface brightness $l_0$
(i.e., luminosity per unit area) convolved with a Gaussian approximated PSF with
covariance $\sigma^2_\Pi$ can be modeled as,
\begin{equation}\label{eq:streakimage}
    m =
    \frac{\ell_0}{2}
    \left[
    \erf\left(\frac{L-2\fcx}{2\sqrt{2\psfwidth}}\right)
    +
    \erf\left(\frac{L+2\fcx}{2\sqrt{2\psfwidth}}\right)
    \right]
    \exp\left(-\half \frac{\fcy^2}{\psfwidth + \sigma_y^2}\right)
    \frac{\sqrt{2\pi\sigma_y^2}}{\sqrt{2\pi(\psfwidth + \sigma_y^2)}}
\end{equation}
where the streak's location, $\left(x_0, y_0\right)$, and position angle,
$\phi_0$, in image space can be described by, 
\begin{align}
    \fc_x &= (x - x_0) \cos\phi_0 - (y - y_0) \sin\phi_0
    \\
    \fc_y &= (x - x_0) \sin\phi_0 + (y - y_0) \cos\phi_0.
\end{align}

For the current consideration of ground-based imaging $\sigma_y^2 \ll \psfwidth$
for this model and can be neglected. Thus our streak model has parameters,
\begin{equation}
    \modelparams \equiv \left[x_0, y_0, \phi_0, \ell_0, L, \psfwidth\right].
\end{equation}
Since we maximize the likelihood with respect to the object flux for detection,
\autoref{eq:streakimage}, we only need to explore the model parameter space,
\begin{equation}
    \modelparamsnoflux \equiv \left[x_0, y_0, \phi_0, L, \psfwidth\right],
\end{equation}
when detecting streaks.


\subsection{Practical Matters}

While in principle one can incorporate the non- flatness and background of the
image into the source model, it is convenient to utilize conventional image
reduction procedures to first produce flat fielded and background subtracted
images and then run the maximum likelihood source detection on these.

It is also convenient when searching for ultra-faint satellites and NEOs to
remove the stars and galaxies from the images before running the maximum
likelihood source detection to detect the satellites and NEOs. There are
numerous ways to do this. One convenient means, when consecutive images have
been taken of the same area of the sky, is median stacking. In the case of
object tracking, a median stack can be created which will effectively filter out
the star streaks, assuming that the density of stars is low enough and exposure
times are short enough that star streaks do not significantly overlap from
exposure to exposure. In the case of sidereal tracking, a median stack can be
constructed which will filter out all objects that move from exposure to
exposure. This median stack can then be subtracted for the individual epochs to
leave images with only the moving objects. This is superior to masking
procedures which often fail to mask the low surface brightness Airy disk wings
of stars that can contribute significantly to false detections when searching
for low surface brightness streaks.

In \autoref{sec:multiobject} we will discuss the practical matter of using the
maximum likelihood method to detect multiple streaks of varying properties in a
single image, as well as deblending.

\section{Demonstrations of the Method}

In this section, we demonstrate the maximum likelihood detection method
discussed in \autoref{sec:method} by applying it to simulations and real data.
We will focus on the more challenging sidereal tracking case (i.e., detecting
streaks). In \autoref{sec:SimSingleStreak}, we apply the method to simulated
data of a single ultra-faint streak showing details of various phases of the
detection pipeline. In \autoref{sec:RealLeo}, we show some of the results of
applying our method in an unsupervised, or blind, fashion to real data
containing a satellite in low earth orbit (LEO). Finally, in
\autoref{sec:multiobject}, we demonstrate how multiple objects can be detected
in a simulated image containing multiple streaks.

\begin{figure}[!htb]
    \centerline{
        \includegraphics[width=0.6\textwidth]{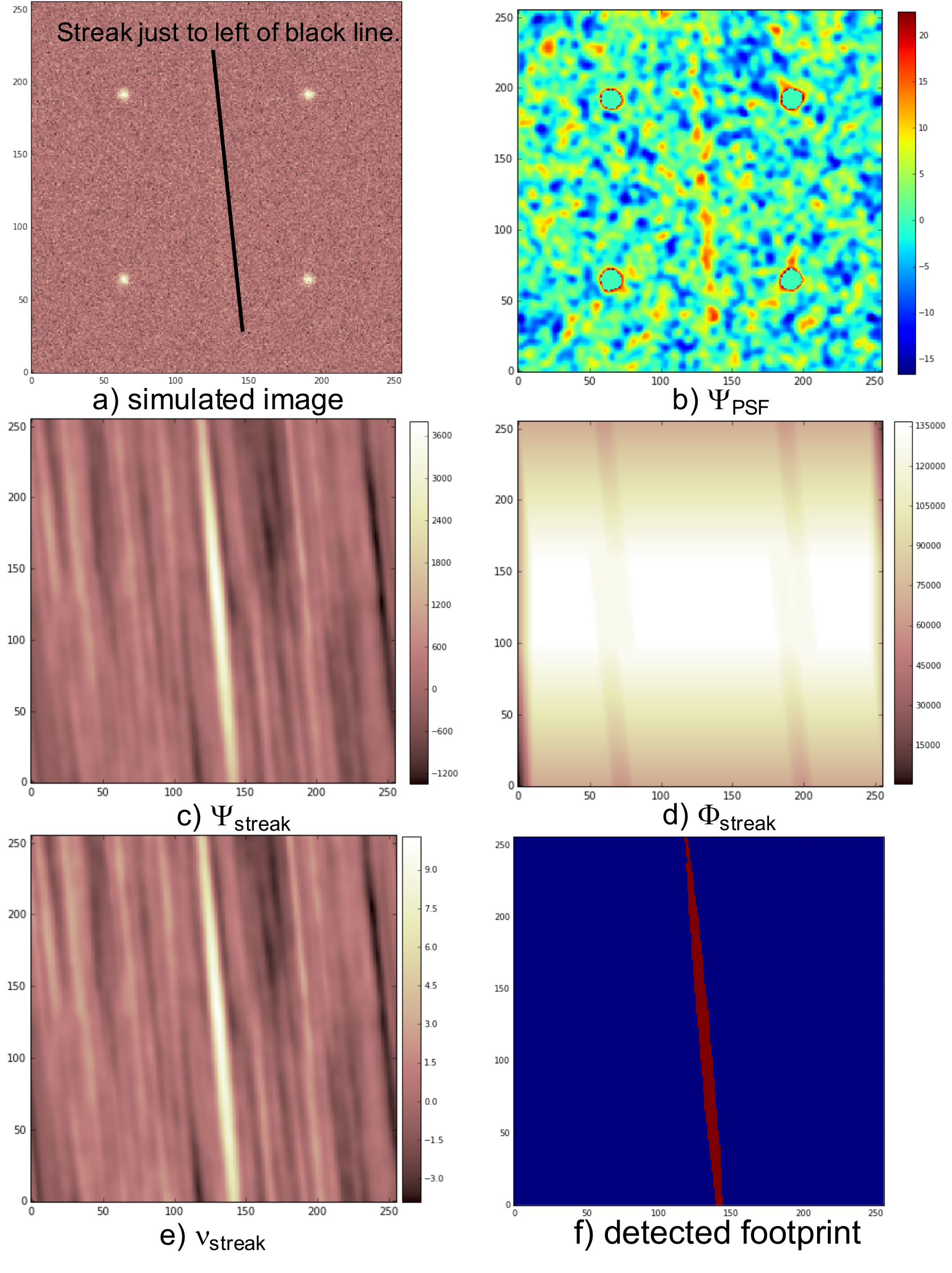}
    }
    \caption{A sequence of images demonstrating various phases of the maximum 
    likelihood reduction pipeline, \autoref{sec:method}.
    \emph{a)} We simulated an ultra-faint streak according to our model,
        \autoref{eq:streakimage}, between four brighter Gaussian PSF stars. The
        black line is an annotation to guide the eye to the simulated streak.
    \emph{b)} We use a Gaussian PSF model in our maximum likelihood formalism to
        detect objects with $\nu_\mathrm{PSF}>5$ and mask these regions from the
        $\Psi_\mathrm{PSF}$ and $\Phi_\mathrm{PSF}$ images. Note that the faint
        streak is just barely visible, but all portions of it are still below
        $5\sigma$ significance.
    \emph{c)} Since the convolutions are transitive we can operate on the
        $\Psi_\mathrm{PSF}$ and $\Phi_\mathrm{PSF}$ images with a simple line
        model (absent the PSF) to generate \emph{c)} $\Psi_\mathrm{streak}$,
        \emph{d)} $\Phi_\mathrm{streak}$, and \emph{e)} $\nu_\mathrm{streak}$
        images for faint streak detection purposes.
    \emph{f)} Shows the faint streak footprint which was detected at more than
        $10\sigma$ significance.
    }
    \label{fig:simsinglestreak}
\end{figure}

\subsection{Simulated Single Ultra-faint Streak}\label{sec:SimSingleStreak}

We simulated a ground-based image containing an ultra-faint streak ($l_0<$RMS
noise) along with four brighter stars, all convolved with a Gaussian PSF, see
\autoref{fig:simsinglestreak}a. We first estimate the PSF, convolve the image
with a Gaussian kernel of that width, and using the maximum likelihood detection
method, detect objects with significance $\nu_\mathrm{PSF} > 5$, where $\nu$ is
defined in \autoref{eq:sig_def}. We then refine our PSF estimate empirically
based on the detected stars and redo the detection process to generate
$\Psi_\mathrm{PSF}$, $\Phi_\mathrm{PSF}$, and $\nu_\mathrm{PSF}$
images\footnote{For example, if the PSF is the same across the image then the
$\Psi_\mathrm{PSF}$, $\Phi_\mathrm{PSF}$ images can be created via fast Fourier
transform (FFT) convolution with the PSF model.}. Using percolation theory to
identify contiguous pixels in the $\nu$ image above the threshold, we then mask
out the pixels associated with star (or bright streaks that were identified
after the PSF convolution) from the various PSF images\footnote{As noted in
\autoref{sec:method} it is often advantageous to median filter/difference rather
than mask.}. For example, we show the masked $\Psi_\mathrm{PSF}$ image in
\autoref{fig:simsinglestreak}b, where hints of the simulated streak is now
apparent but still low significance.

Because convolution operations are transitive we can operate on the PSF
convolved images with a simple line model and it is the same as operating on the
raw data with a PSF convolved streak model, \autoref{eq:streakimage}. By
convolving the PSF convolved images with lines of varying $L$ and $\phi_0$, we
can determine which combination results in the maximum
$\mathrm{ln}\mathcal{L}_\mathrm{ML}$. We performed a grid based
search\footnote{There are many ways of determining the optimal streak model for
detection purposes that are both more efficient and sophisticated. We
demonstrate such a method in \autoref{sec:multiobject}.} in $L$ and $\phi_0$ to
determine the streak model that maximized streak model significance.
\autoref{fig:simsinglestreak}c, d, and e show the intermediate maximum
likelihood products for the optimal streak model. From
\autoref{fig:simsinglestreak}e we see that we are able to detect the ultra-faint
streak at greater than 10 significance. \autoref{fig:simsinglestreak}e shows the
detected footprint that can be used to identify pixels that can be used to
measure the properties of the streak. Alternatively, we can use the optimal
streak model as an estimate of the streak properties.

\subsection{Real LEO Ultra-faint Streak}\label{sec:RealLeo}

\begin{figure}[!htb]
    \centerline{
        \includegraphics[width=0.9\textwidth]{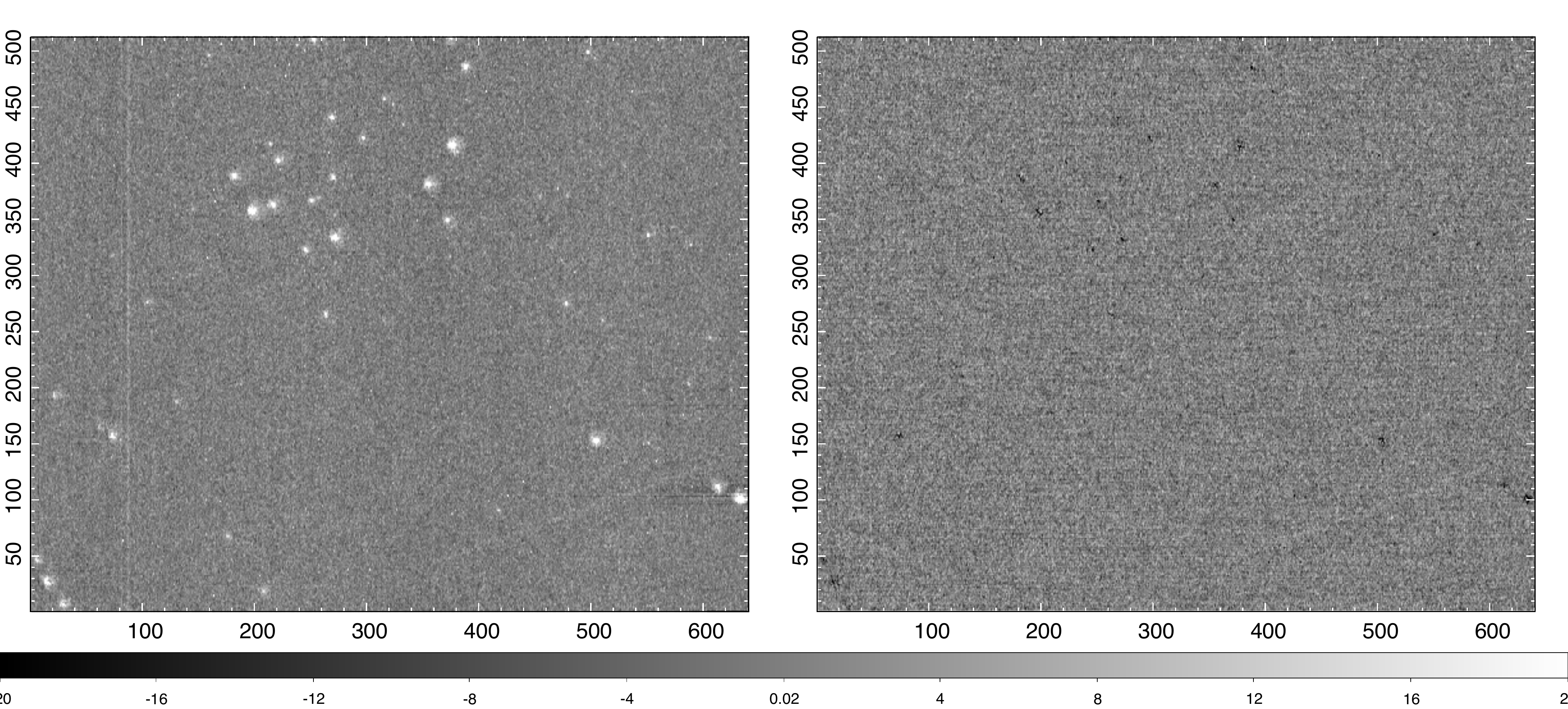}
    }
    \caption{\emph{Left:} An image from the Space-based Telescopes for 
    Actionable Refinement of Ephemeris (STARE) pilot survey \cite{Simms:2012ea, 
    Simms:wz}. The ground-based survey was conducted in sidereal tracking mode 
    thus stars appear as atmospheric PSF convolved point sources, and satellites
    will appear as streaks. \emph{Right:} The same image after subtracting an 
    image of the same field taken one minute earlier. Thus the majority of the 
    stellar flux has been subtracted, however sources that move appreciably over
    the period of one minute (e.g. LEO's) will remain. Operating on the 
    differenced image enables a much higher streak detection purity, with minor
    completeness loss due to slightly increased noise in the differenced image
    (which could further be minimized by differencing a median stacked image 
    rather than a single image). Note that the differenced image also removes 
    static detector defects like the vertical line at the $x\sim90$ pixel 
    location in the left image. Both images are to the same scale. 
    }
    \label{fig:leoimage}
\end{figure}

\begin{figure}[!htb]
    \centerline{
        \includegraphics[width=0.9\textwidth]{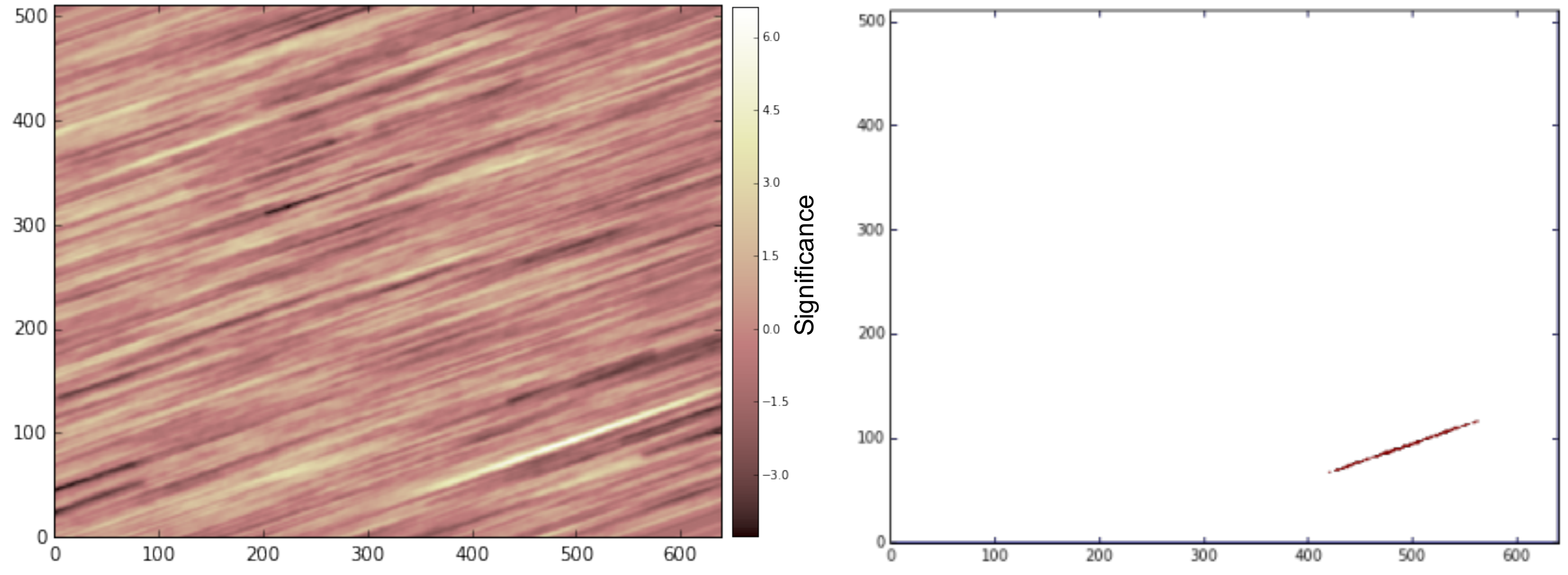}
    }
    \caption{Maximum likelihood detection pipeline products for the images in 
    \autoref{fig:leoimage}. 
    \emph{Left:} The $\nu_\mathrm{streak}$ with the maximum significance after 
        running a grid based model search in $\modelparamsnoflux$ parameter
        space. A streak in the lower right of the image is detected with 
        $>6\sigma$ significance.
    \emph{Right:} The detected streak footprint after applying a $5\sigma$ 
        detection threshold to the $\nu_\mathrm{streak}$ image.
             }
    \label{fig:leodetect}
\end{figure}

In this section we show the results of applying our maximum likelihood source
detection to real data from the Space-based Telescopes for Actionable Refinement
of Ephemeris (STARE) pilot survey \cite{Simms:2012ea, Simms:wz}.
\autoref{fig:leoimage} shows an image from this survey that we analyzed as part
of our automatic detection pipeline, which is similar to that outlined in
\autoref{sec:SimSingleStreak}. One key difference being that we used image
subtraction to remove the stars, rather than masking, before running the maximum
likelihood streak detection method (see the right panel of
\autoref{fig:leoimage}). While no streak was originally detected in the survey,
we were able to detect an ultra-faint LEO object in the lower right corner of
the image with significance $\nu>5$, see \autoref{fig:leodetect}.

\subsection{Detecting Multiple Arbitrary Streaks in Single Image}\label{sec:multiobject}

As demonstrated in the previous sections it is relatively straightforward to
detect a single streak in an image using the maximum likelihood method. However,
there are often multiple streaks in a single image and these may have very
different properties $\modelparamsnoflux$. Ideally any detection method will
simultaneously find all streaks in an image, regardless of varying streak
locations, lengths, and orientations, as well as deblend streaks and maintain a
high purity and completeness (i.e., minimize false detections while maximizing
the number of detected sources). In this section, we show that the maximum
likelihood method derived in \autoref{sec:method} can be formally extended to
enable unsupervised detection of multiple sources.

\begin{figure}
\centering
\begin{subfigure}{3.25in}
  \centering
  \includegraphics[width=3.25in]{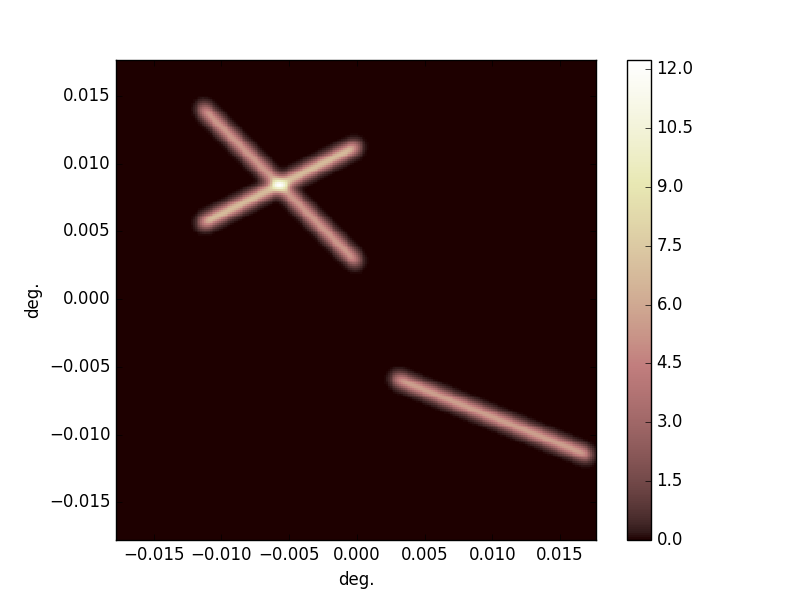}
  \caption{Noise Free}
  \label{fig:3nonoise}
\end{subfigure}%
\begin{subfigure}{3.25in}
  \centering
  \includegraphics[width=3.25in]{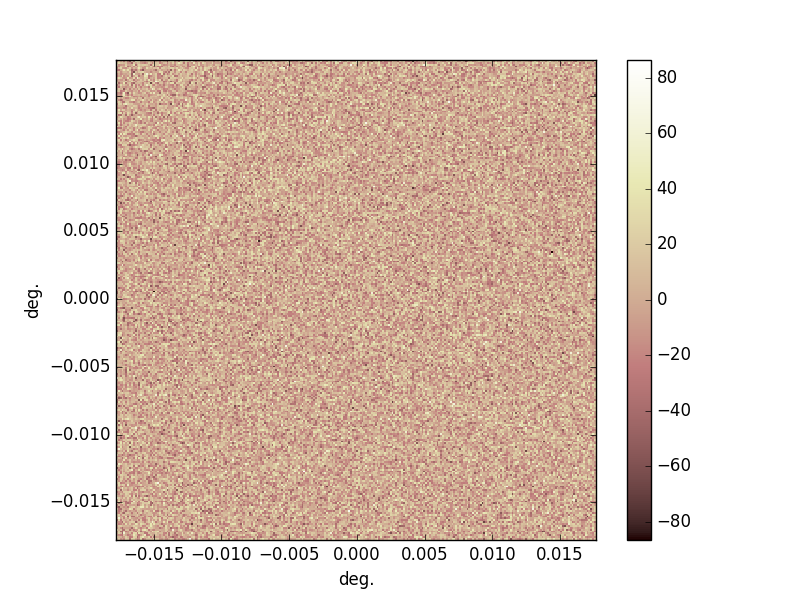}
  \caption{With Noise}
  \label{fig:3withnoise}
\end{subfigure}
    \caption{A simulated image containing three streaks convolved with a 
    Gaussian PSF. Figure \ref{fig:3nonoise} shows the three streaks without any
    noise but including the PSF. Figure \ref{fig:3withnoise} is the same as 
    Figure \ref{fig:3nonoise} but with significant random noise added such that
    the streak surface flux density is $<<$ than the RMS of the pixel noise.
    }
\label{fig:3streaksim}
\end{figure}

To demonstrate this capability, we simulated an image containing three streaks,
Figure \ref{fig:3withnoise}. As can be seen from the noise free version of the
image, Figure \ref{fig:3nonoise}, which is used only for presentation, two of the
streaks have the exact same location parameters, $(x_0, y_0)$, but different
orientations.

The key to detecting multiple streaks in a single image is construction of the
streak model parameter $\modelparamsnoflux$ posterior probability densities for
an entire image\footnote{It is possible to avoid sampling in $\sigma_\Pi$ if
operating on the PSF convolved images.}. This posterior probability density
summarizes the probability of having sources across the range of parameter
space. This posterior can be constructed by using the log likelihood,
\autoref{eq:loglike}, with a sampling method. Some examples of valid sampling
methods include Markov Chain Monte Carlo (MCMC), or random/grid-based searches
coupled with Sequential Importance Resampling (SIR). \cite{Schildknecht2015}
successfully implemented an ad hoc variant of this method by performing a grid
based search in $L$ and $\phi_0$ parameter space, using the ``angle history'',
and ``length history'' as a means of quasi-importance resampling.

\begin{figure}[!htb]
    \centerline{
        \includegraphics[width=0.5\textwidth]{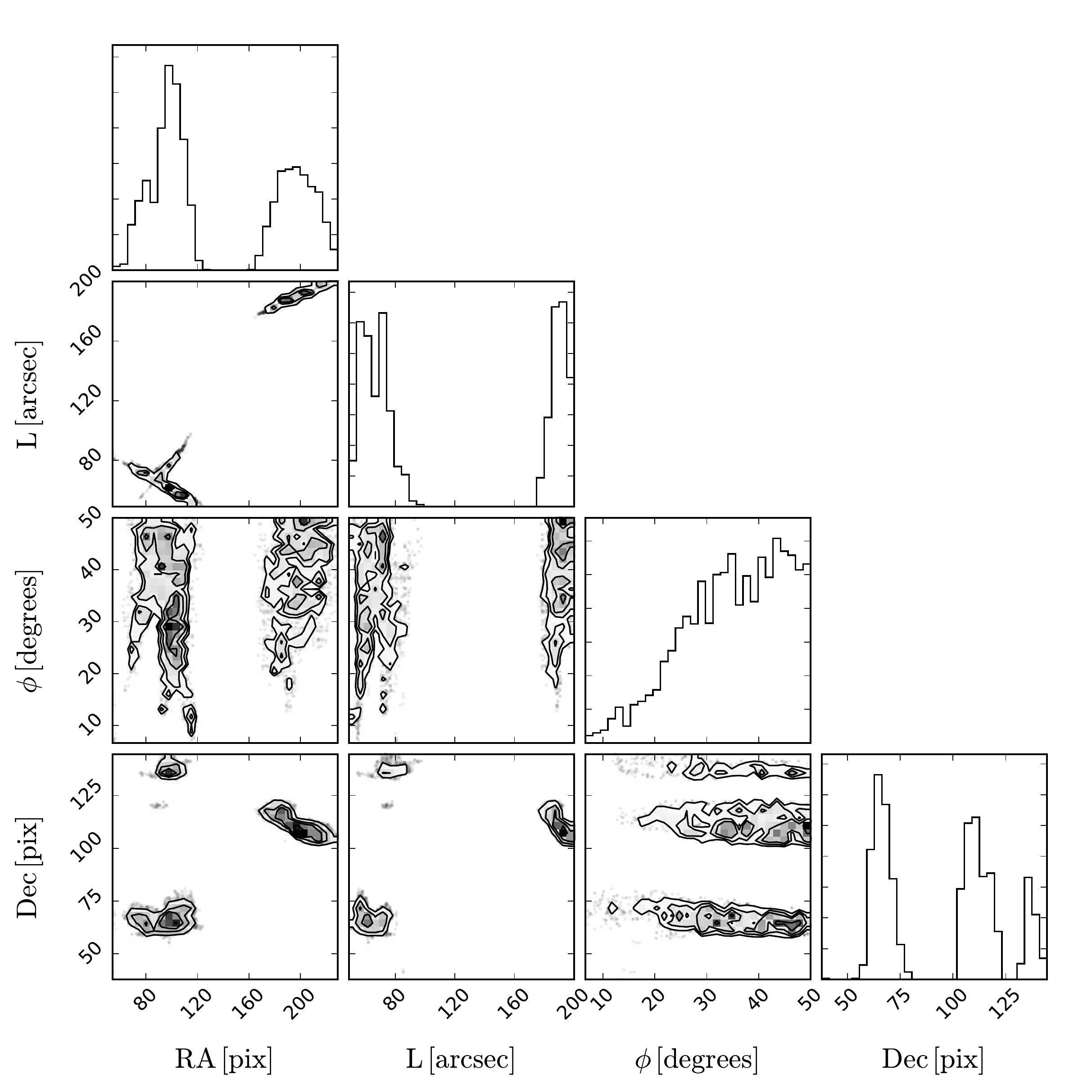}
    }
    \caption{The posterior probability density for the streak model space $\modelparamsnoflux$ resulting from a SIR analysis and using the likelihoods
    from a random search of \autoref{fig:3withnoise} with the maximum likelihood
    detection method. To reduce memory load we only saved samples with
    $\nu_\mathrm{streak}>5$. Note that the Dec.-axis has been flipped relative
    to the y-axis of \autoref{fig:3streaksim}.
    }
    \label{fig:3streakposterior}
\end{figure}

For this work, we randomly sampled 10,000 points in $L$ and $\phi_0$ parameter
space\footnote{The method detects all streaks with as few as 1000 points. We
increased the number purely for the sake of presentation.}, convolving the
$\Psi_\mathrm{PSF}$ and $\Phi+\mathrm{PSF}$ images each time with the given
streak line model. Thus we had $10000\times N_\mathrm{pixels}$ samples. To
reduce memory load we only saved samples with $\nu>5$, however this GMM method
does not require such an arbitrary sample selection. Given the $\nu$ estimates
associated with each sample we calculated the corresponding posterior
probability and used those to resample with a SIR method. The resulting
posterior is shown in \autoref{fig:3streakposterior}.


\begin{figure}[!htb]
    \centerline{
        \includegraphics[width=0.5\textwidth]{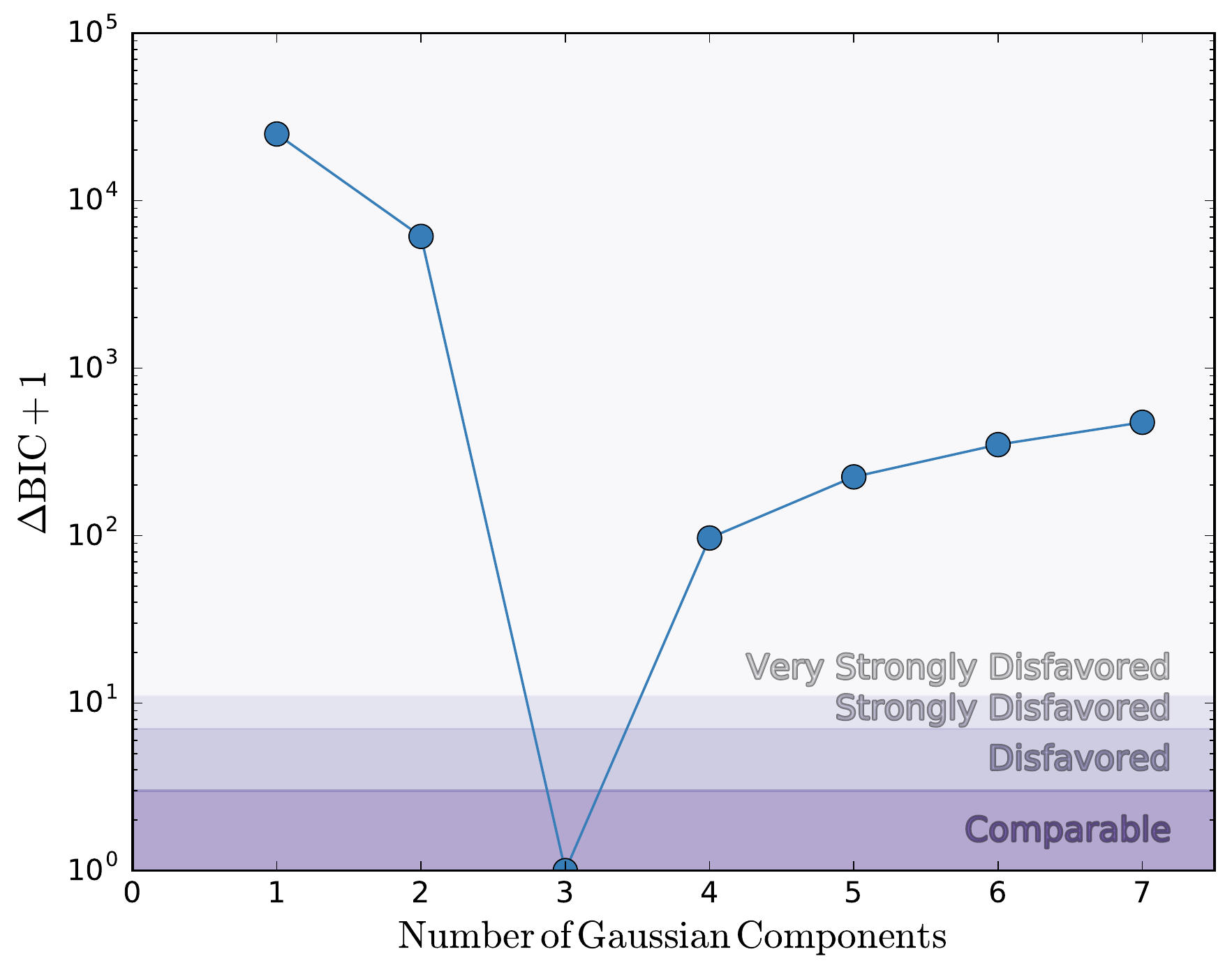}
    }
    \caption{A $\Delta$BIC plot comparing GMM fits to the four-dimensional 
    streak model parameter (right ascension, declination, length, and position
    angle) posterior distribution from an MCMC analysis of Figure
    \ref{fig:3withnoise}., with varying number of Gaussian components and 
    covariance type. The GMM analysis was done assuming a fully variable 
    covariance type. The purple shaded regions roughly denote how a given model
    compares with the model that has the lowest BIC score. The best fit is a
    three component model, meaning that all three simulated streaks were
    detected and there were no false detections.
    }
    \label{fig:delta_bic}
\end{figure}

After constructing the posterior we can use this with unsupervised clustering
algorithms to determine the number and properties of the streaks. In a similar
fashion to the work of \cite{Dawson:2014ve}, we implement \emph{scikit- learn}'s
\cite{scikit-learn} GMM program and apply it to the four- dimensional posterior.
We consider mixtures of 1 to 7 multivariate Gaussian components with fully
unstructured covariance structures.  For each number of components ($n$) we
calculate the Bayesian Information Criterion (BIC) and use this to infer the
optimal number of subclusters. We plot these results as,
\begin{equation}
\Delta\mathrm{BIC}_{n} = \mathrm{BIC}_{n} -  \min\left(\mathrm{BIC}_{n}|n\in\mathbb{Z}_{1...7}\right)
\end{equation}
where $\mathbb{Z}_{1...7}$ is the set of integers from 1 to 7. For convenience
of interpretation we color-code regions of the $\Delta$BIC plot according to the
broad model comparison categories suggested by Kass et al.~\cite{Kass:1995}.
From \autoref{fig:delta_bic} we see that the GMM $\Delta$BIC strongly favors the
model with three streaks.

\begin{figure}[!htb]
    \centerline{
        \includegraphics[width=0.9\textwidth]{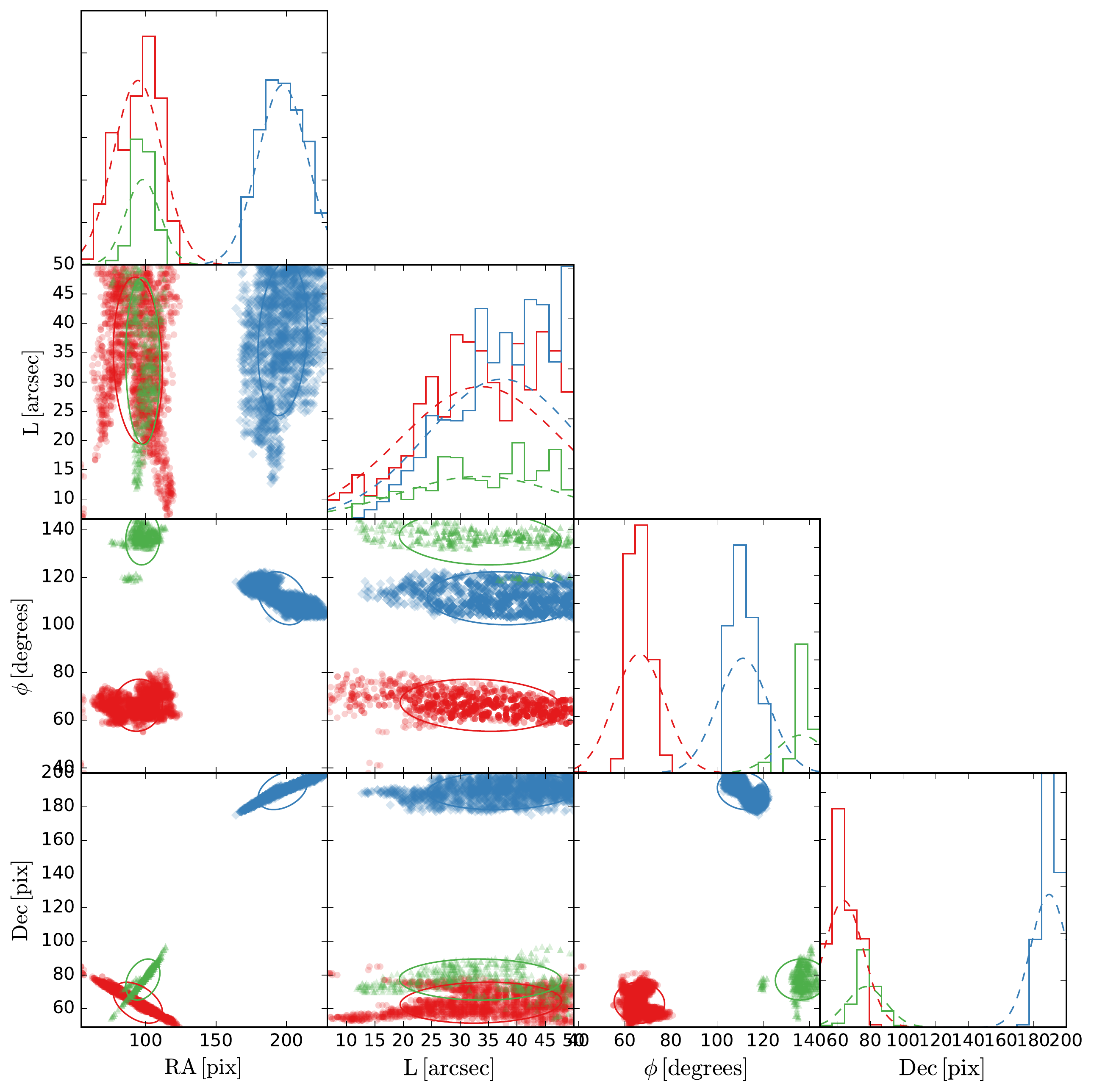}
    }
    \caption{Four-dimensional posterior distribution of the streak model SIR fit
    to Figure \ref{fig:3withnoise} and their most likely cluster membership
    assignment for the best fit GMM (see \autoref{fig:delta_bic}). For the
    projected one-dimensional distributions we plot the marginalized Gaussian
    components for the best fit model (dashed lines). For the projected 
    two-dimensional distributions we plot projected ellipses that encompass
    $\sim68\%$ of the corresponding members in the best fit model Gaussian
    components.  The differently colored ellipses are non-overlapping in many
    2D projections, which is how we are able to effectively deblend the streak
    images.
    }
    \label{fig:gmm_triangle}
\end{figure}

In \autoref{fig:gmm_triangle} we show the parameter posterior distribution but
with samples color coded according to their most likely cluster membership
assignment of the three-component component model. We see that while the
components may overlap in some parameter spaces the fact that we can utilize all
four-dimensions simultaneously we can deblend objects so long as they do not
completely overlap in all parameter dimensions. Furthermore, the clustered
samples provide estimates of the streak properties, including uncertainties, and
can also be used as priors in subsequent forward modeling of the detected
sources (see for example \cite{SchneiderDawson}).

\section{Summary \& Discussion}\label{sec:discussion}

Signal-matched filters have been used extensively in astronomical image
detection algorithms to detect faint, low surface brightness, sources. In this
conference proceeding we have outlined how signal-matched filters can be
incorporated into a maximum likelihood statistical formalism. This formalism has
a number of advantages. It accounts for spatially-varying source models, image
noise, and PSFs. In addition, it enables easy epoch co-addition that is absent
of the systemics associated with varying PSFs and the processing of data on a
pixel grid.

In addition to outlining the maximum likelihood detection method we have
demonstrated it on a number of simulated images, as well as an image of a real
LEO object observed during the STARE proof of concept survey. We have also
demonstrated how the method enables us to detect and deblend multiple
ultra-faint streaks, with varying properties, in a single image.

The ability of the maximum likelihood method to recover all of the information
in a given image, and combine the information from multiple epochs in a
statistically rigorous manner, enables the detection of ultra-faint sources.
Furthermore, the probabilistic basis of the maximum likelihood detection method
enables detection of multiple objects in a given image (including blended
sources), and initial estimates of the detected object properties, including
uncertainties that can be consistently propagated to downstream estimates (e.g.,
orbit parameter estimates; see \cite{SchneiderDawson}). It is for these reasons
that ultra-faint, low signal-to-noise, sources can still result in actionable
information.

\section*{Acknowledgments}

We thank Jim Bosch for introducing us to concept of maximum likelihood source
detection at the Large Synoptic Survey Telescope Dark Energy Science
Collaboration meeting, on February 2, 2015, during his lecture titled ``Image
processing algorithms: Building science-ready catalogs''.  We thank Willem de
Vries for data from the STARE project used to test our algorithms and many
useful discussions. We also thank Mike Pivovaroff, and Alex Pertica for useful
discussions. This work was performed under the auspices of the U.S. Department
of Energy for  Lawrence Livermore National Laboratory under Contract DE-
AC52-07NA27344. Funding for this work was provided by LLNL Laboratory Directed
Research and Development grant 16-ERD-013.

\bibliographystyle{plain}
\bibliography{mldetect}
\end{document}